\documentstyle[psfig,preprint,aps]{revtex}

\tightenlines

\def\beq{\begin{equation}}
\def\eeq{\end{equation}}
\def\bea{\begin{eqnarray}}
\def\eea{\end{eqnarray}}

\begin{document}
\hoffset-1cm
\draft
\title{Gluon Condensate and Non-Perturbative Quark-Photon Vertex
\footnote{Supported by BMBF, GSI Darmstadt, DFG, and Humboldt foundation}}

\author{Munshi G. Mustafa$^1$\footnote{Humboldt fellow and on leave of
absence from Saha Institute of Nuclear Physics, 1/AF Bidhan Nagar, 
Calcutta 700 064, India},
Andreas Sch\"afer$^2$ 
 and Markus H. Thoma$^3$\footnote{Heisenberg fellow}}
\address{$^1$Institut f\"ur Theoretische Physik, Universit\"at Giessen, 35392 
Giessen, Germany}
\address{$^2$Institut f\"ur Theoretische Physik, Universit\"at Regensburg, 
93040 Regensburg, Germany}
\address{$^3$Theory Division, CERN, CH-1211 Geneva 23, Switzerland} 

\date{\today}

\maketitle

\begin{abstract}

We evaluate the quark-photon vertex non-perturbatively
taking into account the gluon condensate
at finite temperature. This vertex 
is related to the previously derived 
effective quark propagator by a QED like Ward-Takahashi 
identity. The importance of the effective vertex 
for the dilepton production rate from a quark-gluon plasma is stressed.
\end{abstract} 

\bigskip

\medskip


\narrowtext
\newpage
During the last two decades a substantial amount of activity, both 
experimentally and theoretically, has been devoted to 
an improved understanding  
of the so called quark-gluon plasma (QGP).
The prime goal is to find a reliable signature for this new phase,
which in turn urgently requires a better
theoretical understanding.
Some insight can be gained within the framework of thermal 
field theory\cite{tft}. Within the perturbative approach the 
thermal effects are known to substantially alter soft modes, 
changing the dispersion relations, 
and allowing for Landau damping as well as
for new modes of propagation for both fermionic and bosonic 
quasi-particles~\cite{klim}. However, perturbative calculations based on 
bare propagators and vertices lead to gauge dependent and infrared 
divergent results.
This inconsistency is due to the fact that at high temperature
higher order loop diagrams can contribute to lower order in the
coupling constant.
This problem has been cured to some extent by the Hard Thermal Loop (HTL)
resummation technique developed by Braaten and Pisarski~\cite{braat}, and much
progress has also been achieved within HTL resummed perturbation theory
in describing dynamical properties of the QGP. In one 
particular application, the dilepton
rate, obtained
by using HTL resummed propagators and vertices~\cite{pisar}, 
was e.g. shown to exhibit
sharp structures in the low invariant mass regime caused by the presence of 
collective quark modes in the medium. However, it seems that 
quite generally non-perturbative 
effects~\cite{boyd}
dominate in the temperature regime attainable in heavy-ion experiments. 

Such non-perturbative effects were made explicit by 
lattice QCD calculations of
{\it e.g.}, effective
parton masses~\cite{pesh}, hadronic correlators~\cite{karsch}, and the gluon 
condensate~\cite{boyd}, below and above the phase transition temperature. 
So far it is beyond the scope of lattice calculations to compute 
dynamical quantities. Gluon and quark condensates, 
however, which describe the
non-perturbative imprints of the QCD ground state, have extensively been 
used in QCD sum rules~\cite{van} for studying hadron properties at zero
and finite temperature. At zero temperature the quark propagator containing
the gluon condensate has been constructed by Lavelle and Schaden~\cite{lave}.
This propagator has been used in a number of investigations~\cite{lave1}.
Recently, as an alternative method to lattice and 
perturbative QCD, it has been suggested in Ref.~\cite{markus} 
to include the temperature-dependent
gluon condensate into parton propagators at finite temperature. 
In this way non-perturbative effects can be taken into account within 
the Greens function technique to compute static~\cite{sch} as well 
as dynamic quantities~\cite{muns}. In the latter work the dilepton 
production rate, which represents a 
promising signature for the QGP formation in 
relativistic 
heavy-ion collisions, has been calculated using an effective quark propagator
containing the gluon condensate. 
Presently such calculations are, however, not yet self-consistent. One has to 
introduce in addition
an effective quark-photon vertex, 
which is related to the effective quark propagator
by the QED Ward-Takahashi identity. As a first step we intend to 
calculate the effective quark-photon vertex in the presence of
the gluon condensate, both at zero and finite temperature, 
and to show that it fulfills
the Ward identity. In a later step we plan to repeat the dilepton calculation 
with this vertex, which is, however, a highly non-trivial 
task in view of the complex form of the result we are going to present. 

The effective vertex, containing  the effects of the gluon condensate
in leading order, is given 
by Fig.~1 and can be written as
\beq
\Gamma^\nu(P_1,P_2) = \Gamma^\nu_0 \ + \ \Gamma^\nu_G(P_1,P_2) \ ,
\label{gamat}
\eeq 
where the bare vertex is given by $\Gamma^\nu_0=-ie\gamma^\nu$ with the
usual QED coupling constant $e$. 
For massless bare quarks the one loop correction
containing the gluon condensate can be written, 
using standard Feynman rules, as
\beq
 \Gamma^\nu_G(P_1,P_2) = - \frac{4}{3} g^2e \int \frac{{\rm d}^4K}{(2\pi)^4}
\ \ \frac{\left [ \gamma^\mu (P_2 \!\!\!\!\!\slash \ + \ K \!\!\!\!\slash \ ) 
\gamma^\nu (P_1 \!\!\!\!\!\slash \ - \ K \!\!\!\!\slash \ ) \gamma^\rho 
\right ]}{(P_1-K)^2 (P_2+K)^2} {\tilde D}_{\mu\rho}(K) \ , \label{gamag}
\eeq
where $g^2=4\pi\alpha_s$ is the strong coupling constant and 
the four momentum is defined as $Q\equiv(q_0,{\vec q})$ and $q=|\vec q|$.
The general expression for the non-perturbative gluon propagator 
containing the gluon condensate is given by~\cite{lave} 
\beq
\tilde D_{\mu\rho}(K) = D^{\rm{full}}_{\mu\rho}(K) - 
D^{\rm{pert}}_{\mu\rho} (K) \ , 
\label{propn}
\eeq
where the subtraction of the perturbative propagator guarantees that
$\tilde D_{\mu\rho}(K)$ is gauge independent. 

The effective quark propagator containing the gluon condensate is given in 
Fig.~2 and can be written as
\beq
S^{-1}(P) = S^{-1}_0(P) - \Sigma_G(P) \ , \label{propqf}
\eeq
where the quark self energy for massless bare quarks reads
\beq
\Sigma_G(P) = \frac{4}{3} \ ig^2 \int \frac{{\rm d}^4K}{(2\pi)^4}
{\tilde D}_{\mu\rho} (K) \gamma^\mu \frac{P\!\!\!\!\slash - K\!\!\!\!\slash}
{(P-K)^2}\gamma^\rho \ \ . \label{selfq}
\eeq
So far we have discussed general expressions for the 
vertex and propagator containing the
gluon condensate. Next we would like to obtain the zero temperature vertex
related to the quark propagator by the Ward-Takahashi identity. 

At zero temperature, transversality
and Lorentz-invariance demand that the gluon propagator has the
following form
\beq
 \tilde D_{\mu\rho}(K)= \left [ g_{\mu\rho} - \frac{K_\mu K_\rho}{K^2} 
\right ] \tilde D(K^2) \ \ . \label{prop0}
\eeq
The lowest moments of the function $\tilde D(K^2)$ of the gluon 
propagator in (\ref{prop0}) are related to the gluon condensate
by~\cite{lave}
\beq
\langle : G^2 : \rangle = \langle :G^{\mu\rho}_a (0)G^a_{\mu\rho}(0):\rangle =
48 i \int \frac{{\rm d}^4K}{(2\pi)^4} K^2
 \tilde D(K^2) \ \ , \label{cond0}
\eeq 
where $a$ is the color index of the SU(3) color gauge group and the gluon field
strengths are summed over color.

Following Ref.\cite{lave} we expand 
the product of the quark propagators in (\ref{gamag}) for small 
loop momentum $K$. Keeping terms which are at most bilinear 
in $K$ leads to
\bea
\frac{1} 
{(P_1-K)^2 (P_2+K)^2}&\approx& \frac{1}{P^2_1 P^2_2} \left [ \ 1 + 2 \frac{K 
\cdot P_1}
{P^2_1} - 2 \frac{K \cdot P_2}{P^2_2} -\frac{K^2}{P^2_1} - \frac{K^2}{P^2_2} +
4 \frac{\left (K \cdot P_1 \right )^2}{P^4_1} \right. \nonumber \\
&&  \left. \hspace{1.2cm} + 4 \frac{\left (K \cdot P_2 \right )^2}{P^4_2} 
- 4 \frac{\left(K \cdot P_1 \right ) \left (K \cdot P_2 \right )}
{P^2_1 P^2_2} \right ] \ .\label{propq}
\eea

Substituting (\ref{prop0}) and (\ref{propq}) into (\ref{gamag}), keeping 
terms only bilinear in $K$, and using (\ref{cond0}) we obtain the following
expression for the one-loop correction
of the vertex:
\bea
\Gamma^\nu_G(P_1,P_2) &=& \frac{ieg^2}{36 P_1^2P^2_2} \left [ \gamma^\nu + 
\frac{4}{3}
(P_1+P_2)^2 \frac{P^\nu_1 P_2 \!\!\!\!\!\slash}{P_1^2P^2_2} 
- \frac{2}{3} \gamma^\nu (P_1+P_2)^2 \frac{P_1 \!\!\!\!\!\slash \  
P_2 \!\!\!\!\!\slash} { P_1^2P^2_2} 
+ \frac{1}{3}\frac{P_1^\nu P_2\!\!\!\!\! \slash}{P_2^2}
+ \frac{1}{3}\frac{P_2^\nu P_1\!\!\!\!\! \slash}{P_1^2} \right. \nonumber \\
&& \hspace{2cm} \left. 
- \frac{5}{3} \frac{P_2^\nu P_2\!\!\!\!\! \slash}{P_2^2}
- \frac{5}{3} \frac{P_1^\nu P_1\!\!\!\!\! \slash}{P_1^2}
\right ] \langle :G^2: \rangle \ \ . \label{gamagf}
\eea
 
Following the same procedure, the quark self energy in (\ref{selfq})
becomes \cite{lave}
\beq
\Sigma_G(P) = g^2 \frac{P \!\!\!\!\slash}{P^4} \ 
\frac{\langle :G^2: \rangle}{36}
\ \ . \label{selfqf}
\eeq

Here we would like to point out that at zero temperature the quark self
energy should also have a contribution containing the quark condensate,
which complicates the problem substantially. 
However, we are interested in the case $T>T_{c}$ and in the deconfining
phase the quark condensate is zero.

Combining (\ref{gamat})
and (\ref{gamagf}) it is now easy to show that the 
effective vertex, containing the
non-perturbative gluon condensate, satisfies indeed the
Ward-Takahashi identity 
\bea
(P_1+P_2)_\nu \Gamma^\nu(P_1,P_2) &=& -ie\left [ \Big ( 
P_1 \!\!\!\!\!\slash \> - \Sigma_G(P_1) \Big ) -
 \Big (-P_2 \!\!\!\!\!\slash \> - \Sigma_G(-P_2) \Big ) \right ] \nonumber \\
 &=& -ie \left [ S^{-1}(P_1) - S^{-1}(-P_2) \right ] \ \ , \label{ward0}
\eea
which relates the vertex to the effective quark propagator, (\ref{propqf})
and (\ref{selfqf}).

In the following we evaluate the vertex at finite temperature.
As usually for
finite temperature a complication arises due to the fact that temperature
is defined in a fixed reference frame and Lorentz invariance is lost. 
Therefore the non-perturbative gluon propagator in (\ref{propn}) 
has a different form for longitudinal and transverse gluons. It
can be written as~\cite{joe} 
\beq
{\tilde D}_{\mu\rho}(K) = P^l_{\mu\rho} {\tilde D}_l(K) +
 P^t_{\mu\rho} {\tilde D}_t(K) \ \ , \label{propr}
\eeq
where the longitudinal and transverse projection tensors are given as
\bea
 P^t_{\mu\rho}+ P^l_{\mu\rho} &=& \frac{K_\mu K_\rho}{K^2} - g_{\mu\rho} 
\nonumber \\
 P^t_{\mu 0} = 0, & \ \ \ & P^t_{ij} = \delta_{ij} - \frac{k_ik_j}{k^2} \ \ .
\label{proj}
\eea

In the rest frame of the heat bath the fermion
self energy for vanishing bare mass can be written as~\cite{weld}
\beq
\Sigma_G(P) = -a(p_0,p) P \!\!\!\! \slash \ - b(p_0,p)\gamma_0 \ \ , 
\label{selfr}
\eeq
with the scalar functions
\bea
 a(p_0,p) & = & \frac{1}{4p^2}\> \left [{\rm {Tr}}\, (P\!\!\!\! \slash \ 
\Sigma_G ) - p_0\> {\rm {Tr}}\, (\gamma _0 \Sigma_G )\right ],\nonumber \\
b(p_0,p) & = & \frac{1}{4p^2}\> \left [P^2\> {\rm{Tr}}\, 
(\gamma _0 \Sigma_G ) - p_0\>
{\rm{Tr}}\, (P\!\!\!\! \slash \ \Sigma_G )\right ]\ ,
\label{scal}
\eea
and $\Sigma_G$ as given by (\ref{selfq}).

Using (\ref{propr}) and (\ref{proj}) in (\ref{gamag}), we get
\beq
\Gamma^\nu_G(P_1,P_2) = \Gamma^\nu_l(P_1,P_2) + \Gamma^\nu_t(P_1,P_2)
\ \ , \label{gamalt}
\eeq
where in the imaginary time formalism
\bea
\Gamma^\nu_l(P_1,P_2) &=& \frac{4}{3} g^2 e\> i T \sum_{k_0}
\int \frac{{\rm d}^3k}{(2\pi)^3}
 \frac{1}{R^2S^2} \left [ \left ( g_{\mu\rho}- \frac{K_\mu K_\rho}{K^2}\right )
 \gamma^\mu R\!\!\!\!\slash \ \gamma^\nu S \!\!\!\!\slash \ \gamma^\rho
- \gamma^i R\!\!\!\!\slash \ \gamma^\nu S \!\!\!\!\slash \ \gamma_i \right.
\nonumber \\
&& \left.\hspace{2cm} -
 \gamma^i R\!\!\!\!\slash \ \gamma^\nu S \!\!\!\!\slash \ \gamma^j 
\frac{k_ik_j}{k^2} \right ] {\tilde D}_l(K^2) , \nonumber \\
\Gamma^\nu_t(P_1,P_2) &=& \frac{4}{3} g^2 e\> i T \sum_{k_0}
\int \frac{{\rm d}^3k}{(2\pi)^3}
 \ \frac{1}{R^2S^2}  \left [ 
 \gamma^i R\!\!\!\!\slash \ \gamma^\nu S \!\!\!\!\slash \ \gamma_i +
 \gamma^i R\!\!\!\!\slash \ \gamma^\nu S \!\!\!\!\slash \ \gamma^j 
 \frac{k_ik_j}{k^2}\right ]
 {\tilde D}_t(K^2) \ , \label{decomlt}
\eea 
with $S=P_1-K$ and $R=P_2+K$. Using again the small $K$ expansion 
(\ref{propq}) and restricting to the lowest
Matsubara mode, $k_0=2\pi i nT=0$, \cite{markus} yields
\bea
 \Gamma^\nu_l(P_1,P_2)&=& \frac{ie}{6}g^2 \frac{1}{P_1^2P^2_2}
\left [ \frac{4}{3}g^{i\nu}\gamma_i -\gamma^\nu + 
\gamma^0 P_2\!\!\!\!\! \slash \ \gamma^\nu P_1\!\!\!\!\! \slash \ \gamma^0
\left ( \frac{1}
{P_1^2} + \frac{1}{P_2^2} + \frac{4p_1^2}{3P_1^4} + \frac{4p_2^2}{3P_2^4}
\right.\right.\nonumber \\
&& \left. \left. - \frac{4({\vec p_1}\cdot {\vec p_2})}{3P_1^2P_2^2} \right )
+ \frac{2}{3P_1^2} \gamma^0 ({\vec p_1}\cdot{\vec \gamma})
\gamma^\nu P_1\!\!\!\!\! \slash \ \gamma^0  
- \frac{2}{3P_2^2} \gamma^0 ({\vec p_2}\cdot {\vec \gamma})
\gamma^\nu P_1\!\!\!\!\!
\slash \ \gamma^0 \right. \nonumber \\
&& \left. - \frac{2}{3P_1^2} \gamma^0 
P_2\!\!\!\!\! \slash \ ({\vec p_1}\cdot {\vec \gamma}) \gamma^0
 + \frac{2}{3P_2^2} \gamma^0 P_2\!\!\!\!\! \slash \ \gamma^\nu 
\ ({\vec p_2}\cdot {\vec \gamma}) \gamma^0 \right ]  \> 
\langle {\cal E}^2\rangle_T\ , \label{gamal}
\eea
and
\bea
 \Gamma^\nu_t(P_1,P_2)&=& -\frac{ie}{12}g^2 \frac{1}{P_1^2P^2_2}
\left [ 2\gamma^\nu - \frac{8}{3}g^{i\nu}\gamma_i+ 
\gamma^i P_2\!\!\!\!\! \slash \ \gamma^\nu P_1\!\!\!\!\! \slash \ \gamma_i
\left ( \frac{2}
{3P_1^2} + \frac{2}{3P_2^2} + \frac{16p_1^2}{15P_1^4} + \frac{16p_2^2}{15P_2^4}
\right. \right.\nonumber \\
&& \left. \left.- \frac{16({\vec p_1} \cdot {\vec p_2})}{15P_1^2P_2^2} 
\right ) 
-\frac{2}{3P_1^2}\gamma^\nu P_1\!\!\!\!\! \slash \ ({\vec p_1}\cdot{\vec 
\gamma}) +\frac{2}{3P_2^2}\gamma^\nu P_1\!\!\!\!\! \slash 
\ ({\vec p_2}\cdot{\vec \gamma})
+\frac{2}{3P_1^2}({\vec p_1}\cdot {\vec \gamma}) 
P_2\!\!\!\!\! \slash \ \gamma^\nu 
\right. \nonumber \\ 
&& \left. -\frac{2}{3P_2^2}({\vec p_2}\cdot {\vec \gamma}) 
P_2\!\!\!\!\! \slash \ \gamma^\nu
 +\frac{2}{3P_1^2}\gamma^i({\vec p_1}\cdot {\vec \gamma}) \gamma^\nu
P_1\!\!\!\!\! \slash \ \gamma_i
 -\frac{2}{3P_2^2}\gamma^i({\vec p_2}\cdot {\vec \gamma}) \gamma^\nu
P_1\!\!\!\!\! \slash \ \gamma_i \right. \nonumber \\
&&\left. - \frac{2}{3P_1^2}\gamma^i P_2\!\!\!\!\! \slash \ 
\gamma^\nu ({\vec p_1}\cdot {\vec \gamma}) \gamma_i +
 \frac{2}{3P_2^2}\gamma^i P_2\!\!\!\!\! \slash \ \gamma^\nu 
({\vec p_2}\cdot {\vec \gamma})
\gamma_i +
\frac{8}{15P_1^4} ({\vec p_1} \cdot {\vec \gamma}) 
P_2\!\!\!\!\! \slash \ \gamma^\nu
P_1\!\!\!\!\! \slash \ ({\vec p_1} \cdot {\vec \gamma}) \right. \nonumber \\
&&\left. +
\frac{8}{15P_2^4} ({\vec p_2} \cdot {\vec \gamma})
 P_2\!\!\!\!\! \slash \ \gamma^\nu
P_1\!\!\!\!\! \slash \ ({\vec p_2} \cdot {\vec \gamma}) -
\frac{4}{15P_1^2P_2^2} ({\vec p_1} \cdot {\vec \gamma}) 
P_2\!\!\!\!\! \slash \ \gamma^\nu
P_1\!\!\!\!\! \slash \ ({\vec p_2} \cdot {\vec \gamma}) \right. \nonumber \\
&& \left. +
\frac{4}{15P_1^2P_2^2} ({\vec p_2} \cdot {\vec \gamma}) 
P_2\!\!\!\!\! \slash \ \gamma^\nu
P_1\!\!\!\!\! \slash \ ({\vec p_1} \cdot {\vec \gamma}) 
\right ]\> \langle {\cal B}^2\rangle_T \ . \label{gamatt}
\eea

At finite temperature the moments of the longitudinal and 
transverse gluon propagator in (\ref{propr}) are related
to the in-medium chromoelectric, $\langle {\cal E}^2\rangle_T$, 
and chromomagnetic condensates, $\langle {\cal B}^2\rangle_T$,
via~\cite{markus}
\begin{eqnarray}
\langle {\bf {\cal E}}^2\rangle _T & = & \langle :G^{0i}_aG_{0i}^a:\rangle _T
= 8T\> \int \frac{{\rm d}^3k}{(2\pi )^3}\> k^2\> \tilde D_l(0,k),\nonumber \\
\langle {\bf {\cal B}}^2\rangle _T & = & \frac{1}{2}\>
\langle :G^{ij}_aG_{ij}^a:\rangle _T
= -16T\> \int \frac{{\rm d}^3k}{(2\pi )^3}\> k^2\> \tilde D_t(0,k).
\label{condt}
\end{eqnarray}

The functions $a$ and $b$ can be expressed
in terms of the chromoelectric and chromomagnetic condensates as~\cite{markus}
\begin{eqnarray}
a(p_0,p) &=& - \frac{g^2}{6} \frac {1}{P^6} \left [ \left ( \frac {1}{3} p^2
- \frac{5}{3}p_0^2\right )\langle {\cal E}^2 \rangle_T - \left ( \frac{1}{5}p^2
- p_0^2 \right ) \langle {\cal B}^2 \rangle_T \right ] , \nonumber \\
b(p_0,p) &=& - \frac{4}{9}g^2 \frac {p_0}{P^6} \left [ p_0^2\langle {\cal E}^2 
\rangle_T 
+ \frac{1}{5}p^2\langle {\cal B}^2 \rangle_T \right ] .  \label{ab}
\end{eqnarray}

Combining (\ref{gamat}), (\ref{gamalt}), (\ref{gamal}), and (\ref{gamatt}) 
and (\ref{propqf}), (\ref{selfr}) and (\ref{ab})
one can show again that the effective vertex, obtained above at finite 
temperature, is related to 
the effective propagator through the Ward-Takahashi identity
\bea
(P_1+P_2)_\nu \Gamma^\nu(P_1,P_2) &=& -ie\Big [  
P_1 \!\!\!\!\!\slash \; \; \Big (1+ a(p_{1}^{0}, p_1) \Big ) \ 
+ b(p_{1}^{0},p_1) \gamma_0 \nonumber \\ &-&
\Big (-P_2 \!\!\!\!\!\slash \; \; \Big (1+ a(-p_{2}^{0}, p_2) \Big ) \ 
+ b(-p_{2}^{0},p_2)\gamma_0 \Big ) 
\Big ] \nonumber \\
 &=& -ie \left [ S^{-1}(P_1) - S^{-1}(-P_2) \right ] \ \ . \label{wardt}
\eea

It can also be shown that, 
keeping terms proportional to $k_0^2$ in 
(\ref{gamal}-\ref{ab}),
taking the zero temperature limit, and replacing $iT\sum_{k_0}
\int{\rm d}^3k/(2\pi)^3
\rightarrow \ \int{\rm d}^4k/(2\pi)^4$ along with $\tilde D_L (k_0,k) =
\tilde D_T(k_0,k) = - \tilde D(K^2) $, the zero temperature
results for the vertex and the propagator are reproduced.

In Minkowski space the condensates in (\ref{condt}) can be expressed 
in terms of the space like ($\Delta_\sigma$) and time like ($\Delta_\tau$) 
plaquette expectation values as ~\cite{markus}
\bea
\frac{\alpha_s}{\pi} \langle {\cal E}^2\rangle_T &=& \frac{4}{11} T^4 
\Delta_\tau
- \frac{2}{11} \langle G^2\rangle_{T=0} \ , \nonumber \\ 
\frac{\alpha_s}{\pi} \langle {\cal B}^2\rangle_T &=& -\frac{4}{11} T^4 
\Delta_\sigma
+ \frac{2}{11} \langle G^2\rangle_{T=0} \ , \label{plaq}
\eea
where $\langle G^2\rangle_{T=0}=(2.5\pm 1.0)T_c^4$~\cite{boyd}.
The plaquette expectation values have been  measured above the critical 
temperature 
$T_c$ on the lattice for  pure SU(3) gauge theory~\cite{boyd}, and
these results can be used to determine the effective vertex and propagator. 
It should be noted that 
our results
are gauge independent but apply only above the critical temperature. 
At zero temperature the quark condensate has to be taken into account,
which generates serious problems with respect to gauge invariance.
The gluon condensate, on the other hand, describing the breaking of 
the scale invariance exists also in the deconfined phase~\cite{leut}.

The effective quark 
propagator gives rise to the quark dispersion relations
discussed in Ref.\cite{markus}. As a 
possible application of the effective Greens functions, containing  
non-perturbative effects, they 
can be used to estimate the dilepton 
production from a QGP. Recently, this rate has
been calculated from the imaginary part of the photon self energy 
in one-loop approximation using our effective quark 
propagator~\cite{muns}. The physical process
corresponding to this rate is the annihilation of collective quark-antiquark 
modes in the QGP. Due to the non-trivial quark dispersion relation
interesting structures (peaks, gaps) show up in the rate, in a similar 
manner as for
the corresponding HTL calculation \cite{pisar}. 
As in the HTL case also the effective
quark-photon vertex, constructed here, should be used in a complete 
calculation of the one-loop photon self energy. 
As a matter of fact, in the case of $\pi^+$-$\pi^-$ 
annihilation, the consideration of the 
Ward identity leads to a strong suppression 
of the dilepton production rate~\cite{korpa}, whereas the use of an effective
HTL vertex results in an enhancement of the QGP rate. Therefore it will be 
very interesting to see, how the gluon condensate affects the balance of both 
effects. The practical calculation seems, however, to be rather involved. 

\bigskip

We gratefully acknowledge the useful discussion with Stefan Leupold and 
Ulrich Mosel during the course of this work.

\begin{figure}
\centerline{\psfig{figure=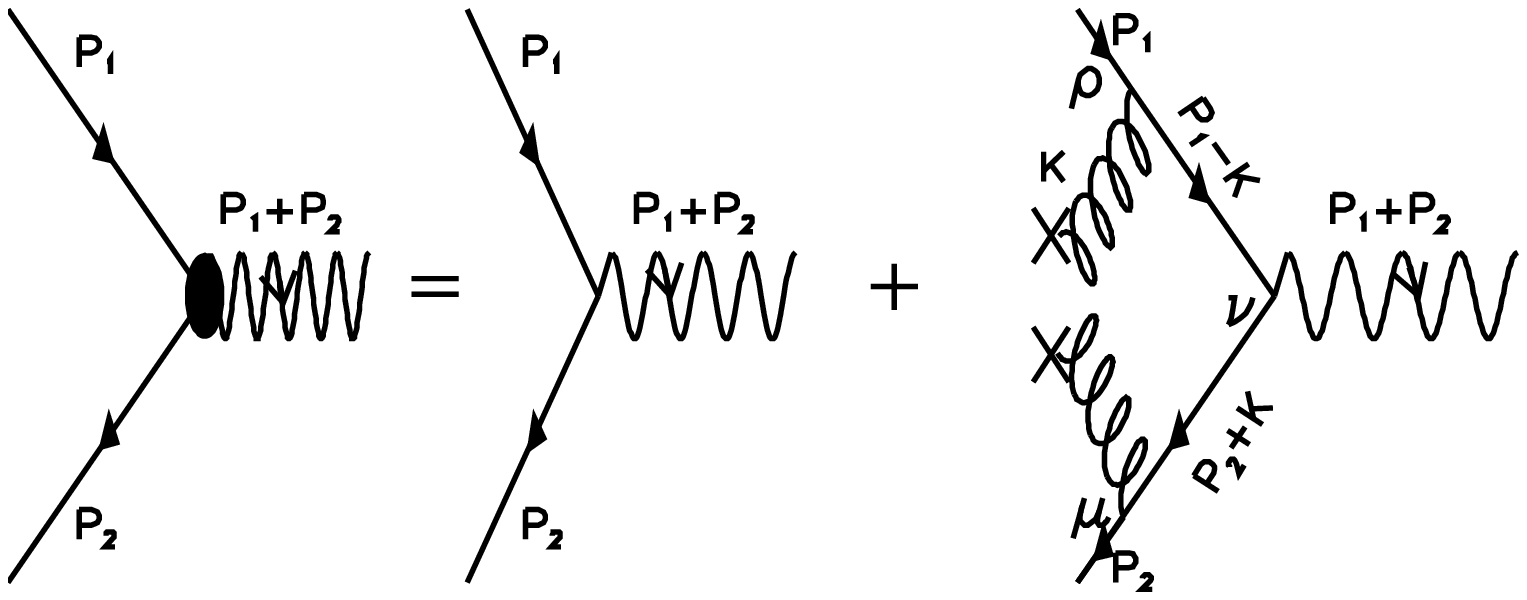,width=14cm,height=16cm}}
\vspace*{-7cm}
\caption{ One-loop correction to the quark-photon vertex containing 
the gluon condensate.}
\end{figure}

\begin{figure}
\vspace*{-3cm}
\centerline{\psfig{figure=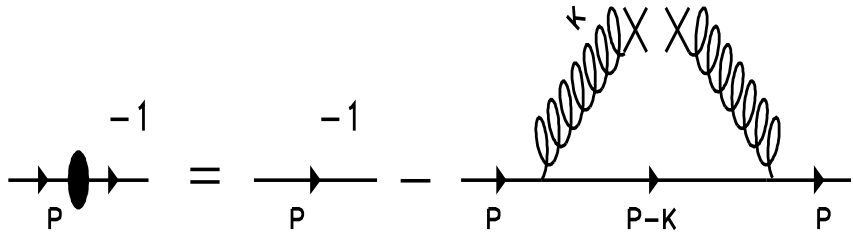,width=14cm,height=18cm}}
\vspace*{-7cm}
\caption{ One-loop correction to the quark propagator containing 
the gluon condensate.}
\end{figure}

\end{document}